\documentstyle[prd,aps,preprint]{revtex}
\begin{document}
\draft

%
%

\preprint{TH-875}
\title{Color Ferromagnetism 
in Quark Matter}
\author{Aiichi Iwazaki}
\address{Department of Physics, Nishogakusha University, Shonan Ohi Chiba
  277-8585,\ Japan.} 
\author{Osamu Morimatsu}
\address{Institute of Particle and  Nuclear Studies, High Energy
  Accelerator Research Organization, 1-1, Ooho, Tsukuba, Ibaraki,
  305-0801, Japan}
\date{July 1, 2003} \maketitle
\begin{abstract}
We show a possibility that there exists a color ferromagnetic state
in quark matter, in which 
a color magnetic field is spontaneously generated.
The state arises between the hadronic state and 
the color superconducting state when the density of quarks is varied.
Although the state ( Savvidy state ) has been known to involve unstable
modes of gluons, we show that the modes compose a quantum Hall state 
to stabilize the ferromagnetic state. 
We also show that the order of the phase transition between the state and
the quark gluon plasma is of the first order. 
\end{abstract}
\hspace*{0.3cm}
\pacs{PACS 12.38.-t, 12.38.MN, 24.85.+p, 73.43.-f  \\ 
Quark Matter,\,\,Color Superconductivity,\,\, Quantum Hall States
\hspace*{1cm}}
\tightenlines
Quark matter possesses several phase structures when its temperature and
densiy are varied; the hadronic phase, the quark gluon plasma phase, and 
the color superconducting phase.
Especially, recent progress\cite{colors} in the color
superconducting phase has been paid much attention. 
The superconducting phase is realized as dynamical effects of quark matter,
namely, the condensation of Cooper pairs of quarks due to
instability of Fermi surface of quark gas against 
attractive forces.
On the other hand, hadronic phase is
realized as dynamical effects of gluons, namely, the condensation
of color magnetic monopoles\cite{monopole}. SU(3) gluons become effectively 
Abelian gluons and color magnetic monopoles at large distance\cite{abelian},
and the monopoles condense to form a dual superconductor.
Therefore, both of these phases are characterized as superconducting
states, i.e. electric and magnetic superconductors.

In this letter we point out a possible existence of a color
ferromagnetic state\cite{savidy} in which the color magnetic field $\vec{B}$ is
spontaneously generated. It is very intriguing that a quantum Hall state\cite{qh} of some gluons, 
which has been known previously to be unstable modes\cite{nielsen}, is
formed to stabilize the ferromagnetic state. Such a quantum Hall state
carrying color charges
is shown to be possible only when quark matter is present. 
The quark matter is shown to have lower energy density
in the ferromagnetic phase than 
that in the color superconducting state when the chemical
potential is less than approximately
$\sqrt{2\langle eB
  \rangle}$. On the other hand, when the chemical potential is 
larger than $\sqrt{2\langle eB \rangle}$, 
the energy density in the superconducting
phase is lower than that in the ferromagnetic state.
Hence, the ferromagnetc phase is expected to arise between the hadronic phase and
the superconducting phase.
The actual value of $\sqrt{\langle eB \rangle}$
is a fairly important quantity characterizing the typical energy scale
of the state although it is 
not yet determined in this paper. 
Here we quote a typical energy scale in QCD, 
$\langle eB \rangle \sim 0.04$
GeV$^2$ 
simply for indicating  a order of the value.
Furthermore,
we show that the ferromagnetic phase of quark matter
may arise at chemical potentials larger than $(3\pi \langle eB \rangle/4L)^{1/3}\sim 180 \,
\mbox{MeV}(\langle eB \rangle /0.04\,\mbox{GeV}^2)^{1/3}(3\,\mbox{fm}/L)^{1/3}$ where
$L$ is the radius of the qurak matter produced in heavy ion collisions.

Let us first explain our strategy. We consider mainly the SU(2) gauge theory 
with massless quarks of two flavours and make a comment on the case of SU(3) gauge theory.
We are concerned with the chemical potential around $500\sim
1000$ MeV so that the coupling is not small. But we
assume that the results obtained with the one loop approximation
are physically correct.
Then, we use the one loop effective potential of color magnetic fields 
calculated previously \cite{savidy}. Similarly, quarks in the magnetic
field are assumed to be almost free particles, which interact with
the spontaneously generated color magnetic field. 
Thus, important quantities in this paper
are the one loop effective potential 
of the color magnetic fields and the energy density of quarks in the field. 
By comparing the energy density of the quarks in the 
color magnetic field with that of the quarks in the 
BCS state, we find that a ferromagnetic phase is realized energetically
for chemical potential less than $\sqrt{2\langle eB \rangle}$.

As is well known\cite{savidy,nielsen}, the effective potential $V$ of the constant color magnetic 
field has the minimum at non-zero magnetic field and also
has an imaginaly part at the minimum:
$V= \frac{11}{48\pi^2}e^2B^2\left(\log
  (eB/\Lambda^2)-\frac{1}{2}\right)-\frac{i}{8\pi}e^2B^2$,
with an appropriate renormarization of the gauge coupling $e$,
where directions of the magnetic field in real space and color space
are arbitrary. In any case of their choices the spontaneous generation of
the magnetic field breaks the spatial rotational symmetry and the color gauge
symmetry. The SU(2) gauge symmetry is broken into the U(1) gauge symmety. 
This fact has made invalid the adoption of the
state as a true vaccum of the gauge
theory. This state, however, may be adopted as a state of the gauge field
in dense quark matter, which is localized for example in
a neutron star or as a nulcear compound object formed by heavy ion collider.  
The possibility is considered seriously in this paper.


The presence of the imaginary part of the potential leads to
excitation of unstable modes around the minimum $eB_{mim}=\Lambda^2$.
This is similar to 
the case that when we expand a potential $=a^2(|\phi|^2-v^2)^2/4$ of a scalar field
around the local maximum, $\phi=0$, i.e. wrong vacuum, unstable modes are present with
their energies $E$ such as $E^2=k^2-a^2v^2<0$. They are excited and
eventually lead to the stable vaccum $\langle \phi \rangle=v$ with condensation of a constant
unstable mode with $E^2(k=0)=-a^2v^2$.
In the gauge theory similar condensations of gluon unstable modes are expected to arise. 


In order to explain the unstable modes in the gauge theory,
we decompose the gauge fields $A_{\mu}^i$ such that
$
A_{\mu}=A_{\mu}^3,\,\,\mbox{and} \,\,
\Phi_{\mu}=(A_{\mu}^1+iA_{\mu}^2)/\sqrt{2}$ 
where indices $1\sim 3$ denote color components. Then, we may suppose 
that the field $A_{\mu}$ is the 'electromagnetic' field 
of the U(1) gauge symmetry and $\Phi_{\mu}$ is the charged vector field,

\begin{eqnarray}
\label{L}
L&=&-\frac{1}{4}\vec{F}_{\mu
  \nu}^2=-\frac{1}{4}(\partial_{\mu}A_{\nu}-\partial_{\nu}A_{\mu})^2-
\frac{1}{2}|D_{\mu}\Phi_{\nu}-D_{\nu}\Phi_{\mu}|^2- \nonumber \\
&-&ie(\partial_{\mu}A_{\nu}-\partial_{\nu}A_{\mu})\Phi_{\mu}^{\dagger}\Phi_{\nu}+\frac{e^2}{4}(\Phi_{\mu}\Phi_{\nu}^{\dagger}-
\Phi_{\nu}\Phi_{\mu}^{\dagger})^2
\end{eqnarray}
with $D_{\mu}=\partial_{\mu}+ieA_{\mu}$,
where we have omitted a gauge term $D_{\mu}\Phi_{\mu}=0$. 
We can derive using the Lagangian that the energy $E$ of
the charged vector field $\Phi_{\mu}\propto e^{iEt}$ 
under the magnetic field $\vec{B}$ is given by
$E^2=k_3^2+2eB(n+1/2)\pm 2eB$  
with a gauge choice, $A_{j}=(0,x_1 B,0)$ and $(\partial_{\mu}+ieA_{\mu})\Phi_{\mu}=0$, 
 where we have taken the spatial direction of $\vec{B}$ being along $x_3$ axis.
$\pm 2eB$ ( the integer $n\geq 0$ ) denote contributions of spin components
of $\Phi_{\mu}$ ( Landau levels )
and $k_3$ denotes
momentum in the direction parallel to the magnetic field.
We note that in each Landau level, there are degenerate
states in momentum $k_2$ with the degeneracy given by 
$eB/2\pi$ per unit area;
$\Phi_{\mu} \sim e^{-ik_2x_2-ik_3x_3}f_n(x_1-k_2/eB)$ where
$f_n(x_1-k_2/eB)$ represents
Harmonic oscillation of the nth order with its center $k_2/eB$ in $x_1$.

It turns out from the spectrum that the states with parallel magnetic moment ($-2eB$) 
in the lowest Landau level ( $n=0$ )
are unstable when $k_3^2<eB$. Therefore, we expect from the lesson
in the scalar field
that the unstable modes are
excited and lead eventually to true stable vacuum with the condensation of
the unstable modes with $k_3=0$. Actualy, there have been several
attempts\cite{nn} 
to find the true vacuum by making the condensation of the unstable modes. 
The difficulty to find the true
vacuum is that since the unstable modes are in the lowest Landau level
so that the condensation of the modes gives rise to 
non-uniform state; their wave functions are localized in
the coordinate $x_1$. Furthermore, it was difficult to see
whether or not any unstable modes dispappear in the condesed state.
In the final section of the paper, we will see that
the excitation of the modes leads to
a uniform quantum Hall state ( QHS ) without any unstable modes. 
Most effecient way to see the uniformness of the QHS is 
using Chern-Simons gauge theory\cite{seme,qh} in $2+1$ dimensions.


First, we estimate energies of quarks in the magnetic field and compare   
energy density of quarks in the field with that of quarks in BCS state.  
First of all, we note that the energy of the quark in the color magnetic field is given by 
$E_{(n,k,s)}=\sqrt{e'B(2n+1-s)+k^2}$, where $e'=e/2$ and
$k$ is a momentum parallel to the magnetic field
and $s$ takes a value of $\pm 1$, representing spin contributions. 
Remember that since we are considering SU(2) gauge
theory with massless quarks of two flavours, 
we have two types of quarks whose color charges are 
positive and negative, respectively for each flavour. 
Both of the quarks have the identical energy. 
There are many degenerate states  
specified by momentum $k_2$ in each Landau level.   
The degeneracy is given by $4\times e'B/2\pi$ per unit area, where the
factor $4$ comes from positive and negative color charged quarks
with each flavour.

The number density of the quarks is estimated such that
$\rho(e'B)=\frac{2e'B}{\pi}\left(\int_{|k|<k_f(n=0)}\frac{dk}{2\pi}+2\sum_{n=1}^{N-1}\int_{|k|<k_f(n)}\frac{dk}{2\pi}\right)=
\frac{e'B}{\pi^2}\left(2\sqrt{\varepsilon_{f}^2}+4\sum_{n=1}^{N-1}\sqrt{\varepsilon_{f}^2-2e'Bn}\right)$,
with $k_f(n)=\sqrt{\varepsilon_{f}^2-2e'Bn}$ and the Fermi energy
$\varepsilon_{f}$ ( $\varepsilon_{f}^2=2e'BN$ ),
where we have assumed for simplicity that all states in the Landau levels $n \leq N-1$ are occupied. 
In the derivation we have used the relation such as
$E_{(n,k,s=1)}=\sqrt{e'B(2n+1-1)+k^2}=E_{(n-1,k,s=-1)}=\sqrt{e'B(2(n-1)+1-(-1))+k^2}$ for
$n\geq 1$. On the other hand, the energy density of the quarks is given
by
$E_{tot}(e'B)=\frac{2e'B}{\pi}\left(\int_{|k|<k_f(n=0)}\frac{dk}{2\pi}\sqrt{k^2}
+2\sum_{n=1}^{N-1}\int_{|k|<k_f(n)}\frac{dk}{2\pi}\sqrt{k^2+2e'Bn}\right)$ 





Then, 
we have proved numerically by equating both densities, $\rho(e'B)=\rho(e'B=0)$
that 
$E_{tot}(e'B)/E_{tot}(e'B=0)< 1 $ 
for any $N$, where the ratio goes smoothly to zero ( one ) as $N$ goes to zero
( infinity ). 
In general, from the dimensional analysis 
the ratio is a function only of the variable, $\rho^{2/3}/e'B$.
Therefore, we find that the energy density of quarks in the magnetic field
with any strength is lower than that of free quarks.

We have estimated numerically how the energy density becomes lower 
with increasing the Fermi energy,

\begin{equation}
\label{ratio2}
\frac{E_{tot}(e'B)-E_{tot}(e'B=0)}{E_{tot}(e'B=0)}=0.266,\quad 0.059,\quad 0.0268,\quad 0.0028
\end{equation}
for $N(=\varepsilon_{f}^2/2e'B) =1$, $2$, $3$, and $10$, respectively.
We have also found numerically that the above quantity goes to zero   
such as $ (E_{tot}(e'B)-E_{tot}(e'B=0)/E_{tot}(e'B=0)\rightarrow
\varepsilon_f^{-4}$ 
with $\varepsilon_f $ increasing.

We should compare these values with those of the energy decrease when
the BCS
state is realized. The energy decrease may be
estimated as follows. That is, only the quarks in  
the vicinity of the Fermi
surface whose width may be given by a gap energy $\Delta$,
gain energy $\Delta$ by making Cooper pairs. Thus the decrease of the 
energy density is given such as $\sim \Delta^2\, \varepsilon_{f}^2$.
Normalizing it by $E_{tot}(e'B=0)$, we find 
$\Delta^2\, \varepsilon_{f}^2/E_{tot}(e'B=0)\propto \Delta^2\,\varepsilon_{f}^{-2}$.
Thus, the decrease of the energy ratio in the BCS state 
is slower than that of the energy ratio in the ferromagnetic state
when the Fermi energy increases.
It implies that the energy gain of quarks in the BCS state is much larger than that in the
ferromagnetic state for sufficiently large chemical potential.
( The one loop effective 
potential energy $V(eB=\Lambda^2)=-11 (eB)^2/96\pi^2$ of the magnetic field 
is much smaller than those
of the quarks so that we have ignored the energy in the discussion. )
Hence, the BCS state is realized for such a high density of quarks.
On the other hand, for sufficiently small chemical potential
the energy gain of quarks in the ferromagnetic state is larger than that 
in the BCS state. Actually, the energy gain \cite{wil} in
the BCS state   
is at most about $4$
percents of $E_{tot}(e'B=0)$ since $\Delta$ has been estimated at most as $50\sim
100$ MeV when the Fermi energy is about $0.5$ GeV.
On the other hand, the energy gain in the presence of
 the magnetic field becomes $6$ percents when
the Fermi energy is equal to $2\sqrt{e'B}$.   
Although we have not yet determined the value of $\sqrt{e'B}$,
we expect that the value of $\sqrt{e'B}$ is order of 
several handred MeV, e.g. $200$ MeV. 
Hence, we find that the phase boundary between the ferromagnetic phase
and the superconducting phase is present around the chemical potential $\mu$ 
of quarks being about $500$ MeV. We are addressed later with the minimum chemical potential
needed for the realization of the ferromagnetic phase.   

Up to now, we have considered the energy densities of the quarks 
and the gauge fields at zero temperature.
It is easy to calculate the free energy at finite temperature of the quarks and the gluons
in the magnetic field. In the case we neglect the contributions of
unstable modes; the modes condense to form a quantum Hall state in which
any unstable modes are absent.



We have numerically estimated the free energy and found that the 
color magnetic field becomes large with the chemical potential. 
This is owing to the magnetic moment generated by the quarks,
$\sqrt{eB}/\Lambda \simeq 1.3,\quad 1.7,\quad 1.9, \quad 2.1,\quad 2.4 
\quad \mbox{for} \quad
\mu/\Lambda=1,\quad 5,\quad 10, \quad 20,\quad 40 \quad
\mbox{respectively}$. 
We note that the strength of
the magnetic field spontaneously generated 
increases very slowly with the chemical potential. 
We have also found that the transition 
from the ferromagnetic field ( $eB\neq 0$ ) to
the quark gluon plasma ( $eB=0$ ) is of the first order.
The critical temperature $T_c$ is given by
$T_c/\Lambda= 0.8,\quad 1.5,  \quad 1.9, \quad 2.1,\quad 2.4 \quad \mbox{for}\quad
\mu/\Lambda=1,\quad 5\quad 10,\quad
20,\quad 40 \quad \mbox{respectively}$.

We proceed to show briefly that the unstable modes compose a quantum Hall state
and that a stable
ferromagnetic state is realized. 
We note that the unstable modes are states in the lowest 
Landau level $n=0$ with the energy given by $E=\sqrt{k_3^2-eB}$;
$k_3^2<eB$. The denenerate states in the Landau level may be specified
by a momentum $k_2$ whose wave functions behave such as
$\exp[-ik_2x_2-ik_3x_3-\frac{1}{2}\,eB(x_1-k_2/eB)^2]$.
In order to form spatially uniform state, 
various modes in $k_2$ with $k_3=0$ must be excited and
we must take average over $k_2$ in a state with the unstable
modes excited. Along this strategy
ones have tried\cite{nn} to find variationally a stable ground state.
However, the resultant state has not been uniform in space 
and not been shown to have no unstable modes
although it is stable in the variational parameters.

In order to see a uniform QHS composed of the unstable modes
we use Chern-Simon gauge theory. Note that the unstable modes with $k_3=0$ 
can be described effectively 
by a $2+1$ dimentional scalar field $\phi=(\Phi_1-i\Phi_2)\sqrt{l/2}$ coupled with
the magnetic field $B=\epsilon_{i,j}\partial_iA_j$
( $\epsilon_{12}=1$ ):
$L=|(i\partial_{\nu}-eA_{\nu})\phi|^2+2eB|\phi|^2-\frac{\lambda}{2} |\phi|^4$, 
with $\lambda=e^2/l$, where $l$ denotes the coherent
length of the magnetic field, namely, 
its extention in the direction of the field and  
the index $\nu$ runs from $0$ to $2$. This Lagrangian
can be obtained by taking only the unstable modes with $k=0$ from eq(\ref{L});
we have ignored the
other modes coupled with the unstable modes.
The second
term represents anomalous magnetic moment which leads to the instability 
of the state $\langle\phi\rangle=0$. The term can be regarded as a negative mass
so that the model is similar to the model of the scalar field with the double
well potential. We see that
the particles of $\phi $ interact with
each other through a delta function potential.
Thus, we expect that the excitations of the unstable modes 
may eventually form a QHS similar to
the case of electrons interacting repulsively with
each others under magnetic field.


In order to describe QHSs, we rewrite the Lagrangian 
by using Chern-Simons gauge field $a_{\nu}$ \cite{seme},

\begin{equation}
\label{la}
L_a=|(i\partial_{\nu}-eA_{\nu}+a_{\nu})\phi_a|^2+2eB|\phi_a|^2-\frac{\lambda}{2}|\phi_a|^4+
\frac{\epsilon^{\mu\nu\lambda}}{4\alpha}a_{\mu}\partial_{\nu}a_{\lambda}
\end{equation}
with antisymmetric tensor $\epsilon_{\mu\nu\lambda}$ 
with $\epsilon^{012}=\epsilon_{12}$, where the statistical factor $\alpha$
should be taken as $\alpha=2\pi\times $integer to keep the
equality of the system described by $L_a$ to that of $L$. Namely, in two dimentional space
the statistics of particles can be changed by attaching a ficticious flux $\int dx^2B_a=2\alpha$
to the particles. The introduction of the field $a_{\mu}$ is 
to attach the flux to the particles of $\phi_a$. 
When we take $\alpha=2\pi\times$ integer, the statistics of the
particles does not change. In our case the bosons of $\phi$ become new bosons
of $\phi_a$. The equivalence of this lagrangian $L_a$ to the original
one $L$ has been shown \cite{seme} in a operator formalism although the
equivalence had been known in the path integral formalism using the world
lines of the $\phi_a$ particles.

It is well known that QHSs can be described by this type
of Chern-Simons gauge theory even in the mean field approximation \cite{qh}.
Equations of motion are given by

\begin{eqnarray}
\label{eq1}
\phi_a^{\dagger}\,i\partial_0\,\phi_a+c.\,c.+2a_0\,|\phi_a|^2=\frac{1}{4\pi}\epsilon_{ij}\,\partial_i\,a_j
\\
\phi_a^{\dagger}\,(\,i\partial_i-eA_i+a_i)\,\phi_a+c.\,c.=\frac{1}{4\pi}\epsilon_{ij}\partial_0\,a_j 
\\
(i\,\partial_0+a_0)^2\,\phi_a-(\,i\vec{\partial}-e\vec{A}+\vec{a}\,)^2\,\phi_a+2eB\,\phi_a=\lambda |\phi_a|^2\phi_a.
\end{eqnarray}
where we have taken $\alpha=2\pi $.

Using these equations,
we now explain how the ferromagnetic state is stabilized by making 
a QHS. The essence is that in the QHS the ficticious flux is
used to cancel on average with the real magnetic flux, i.e. $a_i=A_i$.  
Consequently, the field $\phi_a$ does not feel any gauge field
and the lagrangian eq(\ref{la}) is reduced to one
representing an usual double well potential.
Hence we find a uniform solution, 
$<\phi_a>=v $: $v$ is obtained by solving the above
equations which are reduced to 
$2a_0\,v^2=\frac{1}{4\pi}\epsilon_{ij}\,\partial_i\,a_j=eB/4\pi$
and $a_0^2+2eB=\lambda v^2$.
This state 
is just a QHS of the field $\phi_a$:
We can show\cite{zhang} that the Hall conductivity of the state is
given by $e^2/2\alpha$ .
In the real QHSs, 
Fermi statistics of electrons is changed to
boson statistics by attaching a ficticious flux. The flux
cancels with real magnetic flux so that the bosonized
electrons $\phi_{electron}$ condense at zero momentum,
making QHSs $<\phi_{electron}>\neq 0$. Such a cancellation, namely, the equality $a_i=A_i$,
holds only for a specific value of the filling factor defined
by $2\pi\rho/eB$, where $\rho$ is the number density of
electrons ( in our case, $\rho$ is given by the 
formula in the left hand side of eq(\ref{eq1}),
i.e. the U(1) color charge
density. ) It follows from the constraint equation (\ref{eq1})
that the filling factor satisfies $2\pi\rho/eB=\pi/\alpha$ only 
when the cancellation ( $eA_i=a_i$ ) holds.
Namely, QHSs arise, for example, at the filling 
factor $2\pi\rho/eB=\pi/\alpha=1/3$ with a choice of $\alpha=3\pi$ for real
electrons, while a QHS of gluons arises at $2\pi\rho/eB=\pi/\alpha=1/2$
with a choice of $\alpha=2\pi$.
Therefore, although the state $\phi_a=0$ is unstable due to the 
negative mass, the condensation of $\phi_a$ leads to a stable ground
state $\phi_a=v$ which can occur only at a specific filling factor.
In the similar way to the case of real electrons \cite{zhang},
we can show that any unstable mode
are absent in the QHS; small fluctuations of $\phi_a$ around
$\phi_a=v$ and quasiparticles ( vortex excitaions ) have real finite gaps.  
The situation is quite similar to the model of the scalar field 
with the double well potential.

We should point out that since
the condensate of $\phi_a$ possesses a color charge ( $\rho=eB/4\pi$ ),
the charge must be supplied from somewhere in color neutral system. 
Without the supplier of the color charge, the condensation
can not arise so that the quantum Hall state is not realized.
Thus, the color ferromagnetic state is not stabilized. 
Quark matter is a supplier of the color charge.
Hence, the stable ferromagnetic state
is possible in dense quark matter;
some color charges of quarks are transmitted to 
the condensate. Then, the color charge density of the quark matter
with the radius $L$ 
should be larger than that of the color condensate, $\rho/L=eB/4\pi
L$. Thus,
it follows that the chemical potential, $\mu$,
should be larger than $(3\pi eB/4L)^{1/3}\sim 180 \,
\mbox{MeV}(eB/0.04\,\mbox{GeV}^2)^{1/3}(3\,\mbox{fm}/L)^{1/3}$ in order for the ferromagnetic 
phase to arise in the quark matter.
Since it is a value nessesary for the realization of the phase,
a critical value separating the two phases,
hadronic phase and ferromagnetic phase is larger than it.   


Up to now, we have discussed the ferromagnetic
state of the SU(2) gauge theory with massless quarks of two flavours.
The orientation in color space of the magnetic field generated spontaneously
can be taken arbitrary in the case of the SU(2) gauge theory.
On the other hand, in the SU(3) gauge theory the different choice of
the orientation 
leads to different physical unstable modes\cite{nn};
three unstable modes are present in general but one of the modes
vanishes in a specific  orientation.
The orientation can be determined in  the quark matter
by minimizing the energy density
of the quarks; 
the energy depends on the orientation of
the magnetic field. Namely, when we take $B/|B|=a\lambda_3+b\lambda_8$
( $ a^2+b^2=1$ ), we impose the color neutrality of the quark matter
and minimize its energy. Then,
we can find the orientation of the magnetic field.
The detail will be published in near future.

Finally,
we show that a real observable magnetic field 
is produced by quarks rotating around
the color magnetic field.
If the number of the color positive charged quarks is 
the same as that of the color negative charged quarks,
the total real magnetic moment produced by the quarks vanishes.
But the number of the color positive charged quarks
and that of the color negative charged quarks is different in
the color neutral system due
to the gluon condensation with the color charges in the QHS.
Therefore, the real magnetic moment produced by, for example, up quarks 
does not vanish. Taking $eB$ being several $0.01 \,\mbox{GeV}^2$ and
the radius of quark matter $L$ being several fm, 
we can show that the real magnetic field with strength $10^{14}\sim 10^{15}$ Gauss is
produced in the color ferromagnetic phase of 
the quark matter, 
which may be generated by heavy ion collisions.

\hspace*{1cm}

We would like to express thanks to
Profs. T. Kunihiro, T. Hatsuda and M. Asakawa
for useful discussions. 
One of the authors ( A. I. ) also
expressess thanks to the member of theory 
group in KEK for their hospitality.


\begin{thebibliography}{99}
\bibitem{colors}K. Rajagopal and F. Wilczek, hep-ph/0011333.
\bibitem{monopole}S. Mandelstam, Phys. Lett. 53B 476 (1975).\\
G. tHooft, Nucl. Phys. B190 455 (1981).
\bibitem{abelian}Z.F. Ezawa and A. Iwazaki, Phys. Rev. D25 2681
  (1982).\\
T. Suzuki and I. Yotsuyanagi, Phys. Rev. D42 4257 (1990).
\bibitem{savidy}G.K. Savvidy, Phys. Lett. 71B 133 (1977).\\
H. Pagels, Lecture at Coral Gables, Florida, 1978.
\bibitem{qh}The Quantum Hall Effect, 2nd Ed., edited by R.E. Prange
  and S.M. Girvan ( Springer-Verlag, New York, 1990 ).\\
Quantum Hall Effects, edited by Z.F. Ezawa ( World Scientific ).\\
Z.F. Ezawa, M. Hotta and A. Iwazaki, Phys. Rev. B46  7765 (1992);
Z.F. Ezawa and A. Iwazaki, J. Phys. Soc. Jpn 61 4133 (1990). 
\bibitem{nielsen}N.K. Nielsen and P. Olesen, Nucl. Phys. B144 376 (1978);
Phys. Lett. 79B 304 (1978).
\bibitem{nn}J. Ambijorn, N.K. Nielsen and P. Olesen, Nucl. Phys. B152
  75 (1979).\\
H.B. Nielsen and M. Ninomiya, Nucl. Phys. B156, 1
  (1979).\\
H. B. Nielsen and P. Olesen, Nucl. Phys. B160 330 (1979).
\bibitem{seme}G.W. Semenoff, Phys. Rev. Lett. 61 516 (1988); 
G.W. Semenoff and P. Sodano, Nucl. Phys. B328  753 (1989).
\bibitem{wil}M. Alford, K. Rajagopal and F. Wilczek, Phys. Lett. B422
  247 (1998).
\bibitem{narison}S. Narison, Phys. Lett. B 387 162 (1996)
\bibitem{zhang}S.C. Zhang, H. Hanson and S. Kilvelson          
Phys. Rev. Lett. 62 82 (1989).
\end{thebibliography}
\end{document}